\documentclass[a4paper,11pt]{article}
\pdfoutput=1 

\usepackage{jheppub} 
\usepackage{amsmath}
\usepackage{amssymb}
\usepackage[T1]{fontenc} 
\newcommand{\be}{\begin{equation}}
	\newcommand{\ee}{\end{equation}}
\newcommand{\beq}{\begin{equation}}
	\newcommand{\eeq}{\end{equation}}
\newcommand{\bea}{\begin{eqnarray}}
	\newcommand{\eea}{\end{eqnarray}}

\title{\boldmath Quantum tunneling from accelerating three-dimensional black hole}

\author[a]{Usman A. Gillani,}
\author[a,1]{Jamil Ahmed, \note{Corresponding author.}}
\author[b]{ and Mudassar Rehman}

\affiliation[a]{National University of Technology, Islamabad, Pakistan}
\affiliation[b]{Department of Mathematics, Quaid-i-Azam University, Islamabad, Pakistan}

\emailAdd{ugilani@nutech.edu.pk}
\emailAdd{jamilahmed@nutech.edu.pk}
\emailAdd{mrehman@math.qau.edu.pk}

\abstract{Hawking radiation as a quantum tunneling phenomena from accelerating BTZ black holes is presented in this work. We have calculated the Dirac particle's Hawking radiation from the horizon of the accelerating BTZ black hole. WKB approximation is applied to Dirac equation in the background of three dimensional black holes. This procedure gives us the tunneling probability, which we have used to calculate the Hawking temperature of the background three-dimensional black hole. Our study is consistent with the previous studies in the absence of the acceleration parameter. We have also studied the quantum corrections to the Hawking temperature of accelerating BTZ black hole. }
\makeatletter
\gdef\@fpheader{}
\makeatother
\begin{document}
\maketitle
\flushbottom

\section{Introduction}

Banados-Teitelboim-Zanelli have obtained the first black hole solution in the (2+1) dimensions with negative cosmological constant, called the BTZ black hole \cite{1,2}. BTZ black holes are particularly important as they provide insight into the physics of black holes, quantum description of gravity and their relation to string theory \cite{3,4,5}. BTZ black holes can be considered as a first model for general AdS/CFT correspondence because these are locally $AdS_{3}$ without curvature singularity \cite{6,7}. Low dimensional models are important in this area of research as one can easily understand the complex problems, such as, loss of evaporation and quantum black holes in these dimensions and then one can readily generalize the results to four and dimensions higher than four. Another important feature of the BTZ black holes is that they can be related to the four and five-dimensional  stringy black holes \cite{8,9,10} which are asymptotically flat.

Black hole solutions related to acceleration are extracted from the C-metric \cite{11,12,13,14}. Peculiar thing about accelerating black holes is that the acceleration comes from the deficit angle of cone in the direction of one polar axis which is responsible for the force. Accelerating black holes have the Hawking temperature which is more than Unruh temperature of the accelerated frame \cite{15}. Thermodynamics and phase transitions of accelerating black holes were studied in Refs. \cite{16,17,18,19,20,21}. Charged accelerating black holes represents different rate of complexity in the late-time growth then the ordinary charged black holes \cite{22}. Work of Refs. \cite{23,24} revealed that the acceleration parameter influence the efficiencies of heat engines. Also the presence of acceleration parameter alter the orbits of the photons which are circular deviated from the equatorial plane. Also there is alteration in the property of the shadow of the black hole \cite{25,26}. Accelerating BTZ black holes in three-dimensions were proposed \cite{26a, 26aa, 27}.

During the process of radiation the loss of information of black hole's radiation is based on two facts. Studies reveal that the black hole's emission spectrum is completely thermal and validity of the no-hair theorem remains intact \cite{27a,27aa}. Thermal spectrum can be fully determined by the temperature, which implies that there is no trace of information in the the outgoing radiation. Also the outside geometry of the black hole can be fully specified by the only parameters  namely, the mass, the charge and the angular momentum of the black hole. This is due to the fact that the no-hair theorem is valid. We can conclude that the spacetime geometry representing the outside of the black hole does not have any information as well. From these two facts, one can conclude that the collapsed matter does not provide any trace of the information. This conclusion is in full agreement with the predictions of the quantum mechanics. On the other hand some information can be gathered if one is able to correlate some of the features of the collapsed matter. This is possible if both the thermal emission and no-hair theorem are not considered simultaneously. Conservation of the energy will be lost in the case,  if any one of these conditions is fully satisfied. Hence to avoid this, the background geometry of the black hole is considered to be fixed and during the emission process from the black hole the conservation of energy condition is enforced. In this regard the work of Ref.  \cite{27a} showed that radiation from the black hole is completely non-thermal and the conservation of energy condition is satisfied, if one considers the dynamical black hole's geometry and the emission as tunneling process.

Using the knowledge of quantum field theory in the black hole's spacetime, Hawking \cite{28} discovered that the black holes could radiate. This study relates general theory of relativity and quantum field theory, it was believed that, deeper understanding of Hawking radiation will give insight to the quantum nature of gravity. The original derivation proposed by Hawking was complicated to generalize for other black holes. Therefore a semi-classical derivation for Hawking radiation as tunneling process was developed \cite{29}. In this approach, the null geodesic equation is used to calculate the imaginary part of the action, from which tunneling probability and hence Hawking temperature can be obtained \cite{30,31,32,33,34,35,36,36a}.

In this work, we have used the method given in Ref. \cite{37} to calculate the Hawking radiation of Dirac particles from accelerating BTZ black hole. One can find extensive studies of Hawking radiation for three dimensional black holes \cite{38,39,40,41}. We have extended this work for three dimensional accelerating black hole.

Through the modified fundamental commutation relation $[x_{i}, p_{j}=i \hbar \delta_{ij}[1+\beta p^{2}]$, one can predict the minimum measurable length as suggested by various theories of quantum gravity. This minimum length can be approached from generalized uncertainty principle (GUP). The expression of GUP is derived as $\Delta x \Delta p \geq \hbar/2[1+\beta (\Delta p)^2]$, where $\beta=\beta_{0}/M_{p}^{2}$. $M_{p}$ is the Planck mass and $\beta$ is a dimensionless parameter \cite{37a}. Quantum properties of gravity can be explored by applying these modifications \cite{37aa, 37aaa, 37aaaa}. Black holes are effective modes to explore the effects of quantum gravity. Therefore we have also studied the quantum corrected Hawking temperature of accelerating BTZ black hole. 

Rest of the work is organised as follows: in Section 2, accelerating BTZ black holes \cite{27} will be presented, then we will present the derivation of Dirac particle's Hawking radiation in Section 3 and investigate the Hawking temperature for these black holes. Quantum corrections to the Hawking temperature has been investigated in Section 4. In the last section we will summarize the findings of our study.

\section{Accelerating BTZ black holes}

\label{sec_ bh}

The Einstein action in (2+1) dimensions in absence of matter takes the following form
\begin{equation}
\mathcal{S}=\frac{1}{16\pi }\int d^{3}x\sqrt{-g}\left( R-2\Lambda \right) ,  \label{E}
\end{equation}%
where $R$ represents the Ricci scalar, $\Lambda$, is used for Cosmological constant and $g$ is for the determinant of the
metric tensor $g_{\mu \nu }$.
Variation of the above action with respect to the metric $g^{\mu \nu}$, gives the following form of the field equations
\begin{equation}
G_{\mu \nu}+\Lambda g_{\mu \nu}=0,
\end{equation}
where $G_{\mu \nu}$ is the Einstein tensor.
The corresponding solution of the above field equations will have the form as proposed in Ref. \cite{27}, representing the accelerating BTZ black hole
\begin{equation}
ds^{2}=\frac{1}{\chi^2}\left[-f\left( r\right) dt^{2}+\frac{1}{f\left( r\right)} dr^{2}+r^{2}
d\phi^{2}\right],  \label{BTZ_}
\end{equation}%
where%
\begin{eqnarray}
\chi\left(r, \phi \right)&=& \alpha r \cosh(\sqrt{m-k} \phi), \\
f\left( r\right) &=& \left(\left(M-k\right)\alpha^2-\Lambda \right)r^2+k -M.
\label{3.13}
\end{eqnarray}
Here $M$ is the mass, $\alpha$ is used to represents the acceleration parameter and $k$ is the topological constant. The spacetime coordinates have the following values
\begin{equation}
0\leq \phi <2\pi, \ \  0\leq r<\infty,\ \  -\infty <t<\infty.
\end{equation}%
The horizons of the line element (\ref{BTZ_}) are given as
\begin{equation}
r_{\pm}=\pm\sqrt{\frac{M-k}{(M-k)\alpha^2-\Lambda}},
\label{3.12}
\end{equation}
where `+' and `$-$' signs denote the locations of the outer and inner horizons.

\section{Hawking temperature of Dirac particles}

\label{sec_ uncharge_fermionic_charge}

In this section we will investigate the Hawking temperature of Dirac particles from the accelerating
BTZ black hole. We consider the massive spinor field $\psi $, which obeys the following covariant form of the Dirac equation
\begin{equation}
i\hbar \gamma ^{a}e_{a}^{\mu } \nabla _{\mu }\psi
-m \psi =0,
\end{equation}%
where $\nabla _{\mu }$ is the covariant derivative of the spinor field given by $\nabla
_{\mu }=\partial _{\mu }+\frac{1}{4}\omega^{ab}_{\mu}\gamma_{[a}\gamma_{b]}$,
with%
\begin{eqnarray}
\omega^{ab}_{\mu} &=&e^{a}_{\nu}\Gamma^{\nu}_{\sigma\mu}e^{\sigma b}-e^{\nu b}\partial_{\mu}e^{a}_{\nu}.
\end{eqnarray}%
In three dimensions the $\gamma $ matrices will have the following form%
\begin{equation}
\gamma ^{b}=\left( i\sigma ^{1},\sigma ^{0},\sigma ^{2}\right) ,
\end{equation}%
where $\sigma ^{i}$ represents the matrices called the Pauli sigma matrices. For the spacetime (\ref%
{BTZ_}) the frame field $e_{a}^{\mu }$ have the following form
\begin{eqnarray}
e_{0}^{\mu } &=&\left(
\begin{array}{ccc}
\chi f^{-1/2} & 0 & 0%
\end{array}%
\right) ,  \nonumber \\
e_{1}^{\mu } &=&\left(
\begin{array}{ccc}
0 & \chi f^{1/2} & 0%
\end{array}%
\right) ,  \nonumber \\
e_{2}^{\mu } &=&\left(
\begin{array}{ccc}
0 & 0 & \frac{\chi}{ r}%
\end{array}%
\right) .
\end{eqnarray}%
Spinor field  $\psi $ can be written in the following form
\begin{equation}
\psi =\left(
\begin{array}{c}
A\left( t,r,\phi \right)  \\
B\left( t,r,\phi \right)
\end{array}%
\right) e^{\frac{i}{\hslash }\mathcal{Y}\left( t,r ,\phi \right) }.
\end{equation}%
Applying the WKB approximation, for which we will use the above form of the spin field into the Dirac equation and after some simplification we get the following system of coupled differential equations
\begin{equation}
A\left( m +\frac{\chi}{ r}\partial _{\phi }\mathcal{Y}\left( t,r,\phi \right) \right) +B%
\left[ \chi\sqrt{f}\partial _{r}\mathcal{Y}\left( t,r,\phi \right) + \frac{\chi}{\sqrt{f%
}}\partial _{t}\mathcal{Y}\left( t,r,\phi \right)
 \right] =0,
\end{equation}%
\begin{equation}
A\left[ \chi\sqrt{f}\partial _{r}\mathcal{Y}\left( t,r,\phi \right) - \frac{\chi}{\sqrt{%
f}}\partial _{t}\mathcal{Y}\left( t,r,\phi \right)
 \right] +B\left( m -\frac{\chi}{r}\partial _{\phi
}\mathcal{Y}\left( t,r,\phi \right) \right) =0.
\end{equation}%
Note that although $A$ and $B$ are not constant, their derivatives are all of order $\hbar ,$ in the lowest order of WKB approximation we neglect the derivatives of $A$ and $B$. The above system of homogenous equations will have a non-trivial solution only if we impose the following constraint
\begin{equation}
\frac{\chi^{2}}{r^{2}}\left( \partial _{\phi }\mathcal{Y}\left( t,r,\phi \right)
\right) ^{2}-m^{2}+\left( \sqrt{\chi^2f}\partial _{r}\mathcal{Y}\left( t,r,\phi
\right) \right) ^{2}-\left( \frac{\chi}{\sqrt{f}}\partial _{t}\mathcal{Y}\left(
t,r,\phi \right)
 \right) ^{2}=0.  \label{wave__}
\end{equation}

In the present case of accelerating BTZ black hole, we have two Killing vectors $\left( \frac{\partial }{\partial t}%
\right) ^{\mu }$ and $\left( \frac{\partial }{\partial \phi }\right) ^{\mu }$, this allow us to use the separation of variable method in the following form
\begin{equation}\label{2a}
\mathcal{Y}=-\omega t+j\phi +W\left( r\right) +K,
\end{equation}%
where $K$ is a complex number and Dirac particle's energy and angular momentum is denoted by $\omega$ and $j$ respectively. Now putting
\begin{equation}
\partial _{r}\mathcal{Y}=\partial _{r}W\left( r\right) ,~~~\partial _{\phi
}\mathcal{Y}=j,~~~\partial _{t}\mathcal{Y}=-\omega ,
\end{equation}%
in (\ref{wave__}) we get
\begin{equation}
\partial _{r}W\left( r\right) =\pm \frac{1}{f}\sqrt{ \omega
  ^{2}-f\left(\frac{j^{2}}{r^{2}}-\frac{m^{2}}{\chi^{2}}%
\right) }.
\end{equation}%
Integration gives
\begin{equation}
W_{\pm}\left( r\right) =\pm\int \frac{dr}{f}\sqrt{ \omega
 ^{2}-f\left( \frac{j^{2}}{r^{2}}-\frac{m^{2}}{\chi^{2}}\right) }%
.  \label{phase_r__}
\end{equation}%
Using residue integration, we will have the tunneling probability in the following form
\begin{equation}
W_{\pm}=\pm\frac{i \pi\omega}{f^{\prime }(r_{+})},
\end{equation}%
which gives
\begin{equation}
\mathrm{Im}W_{\pm}=\frac{\pi \omega}{2\kappa  },
\end{equation}%
where
\begin{equation}
2\kappa=f^{\prime }(r_{+})=2r_{+}\left( \left(M-k\right)\alpha^2-\Lambda\right).
\end{equation}
The process which is forbidden classically can be done by calculating the imaginary part of the action given above. In order to calculate the probability of Dirac particles across horizon we use this semi-classical approach, which yields the following \cite{37}
\begin{eqnarray}
P_{Out} &\varpropto &\exp \left( \frac{-2}{\hbar }\mathrm{Im}\mathcal{Y}\right)
=\exp \left( \frac{-2}{\hbar }(\mathrm{Im}W_{+}+\mathrm{Im}K)\right) ,
\label{p_emit_} \\
P_{In} &\varpropto &\exp \left( \frac{-2}{\hbar }\mathrm{Im}\mathcal{Y}\right)
=\exp \left( \frac{-2}{\hbar }(\mathrm{Im}W_{-}+\mathrm{Im}K)\right) .
\label{p_absorb_}
\end{eqnarray}%
Thus we can write the probability of the Dirac particles across the horizon as
\begin{equation}
\Gamma =  \frac{P_{Out}}{P_{In} },
\end{equation}%
which gives
\begin{equation}
\Gamma =\exp \left( \frac{-4}{\hbar }\mathrm{Im}W_{+}\right) ,
\label{rate_phase_}
\end{equation}%
using the value of $W_{+}$, we have
\begin{equation}
\Gamma =\exp \left[ -\frac{2\pi \omega}{\hbar \kappa }
\right] .
\end{equation}%
Boltzmann factor is given by $\Gamma =\exp\left( -\beta \omega \right)$, here  $\omega$ is the energy of the particle and $\beta$ is the inverse of Hawking temperature  \cite{37}. Now taking $\hbar=1$, we get the following analytical expression for Hawking temperature
\begin{equation}  \label{1}
T_{H}=\frac{r_{+}\left( \left(M-k\right)\alpha^2-\Lambda\right)}{2\pi}.
\end{equation}
This result shows that the Hawking temperature for accelerating BTZ black hole is dependent on mass and the acceleration parameter.

\section{Quantum corrections to Hawking temperature}

In this section we will consider the effects of quantum gravity in the tunneling process of Dirac particles across the event horizon of the accelerating BTZ black holes. For this purpose we will consider the generalized Dirac equation which is given by \cite{dirac}
\begin{eqnarray}\label{q1}
\bigg\{i \gamma^{0} \partial_{0}+i \gamma^{k}(1-\beta m)^{2} \partial_{k}+i \gamma ^{k} \beta \hbar^{2}(\partial_{j} \partial^{j}) \partial_{k}+\frac{m}{\hbar}(1-\beta m^{2} +\beta \hbar^{2} \partial_{j} \partial^{j})  \nonumber \\
+i \gamma^{mu} \omega_{\mu} (1-\beta m^{2} +\beta \hbar^{2} \partial_{j} \partial^{j})\bigg\}\psi =0, 
\end{eqnarray}
where $\beta$ is a small quantity which belongs to the effects of quantum gravity. The Latin indices belong to flat metric $\eta_{\mu\nu}$ while the Greek indices live in the curved metric $g_{\mu\nu}$. In the above equation
\begin{eqnarray}
\gamma^{a}&=&\bigg(\left(
                     \begin{array}{cc}
                       0 & 1 \\
                       -1 & 0 \\
                     \end{array}
                   \right),
                   \left(
  \begin{array}{cc}
    0 & 1 \\
    1 & 0 \\
  \end{array}
\right),
\left(
  \begin{array}{cc}
    1 & 0 \\
    0 & -1 \\
  \end{array}
\right)
\bigg), \nonumber\\
\gamma^{\mu}&=&e^{\mu}_{a}\gamma^{a},\nonumber\\
e^{\mu}_{a}&=&diag\bigg(\frac{\chi}{\sqrt{f}}, \chi\sqrt{f},\frac{\chi}{r}\bigg),\nonumber\\
\omega_{\mu}&=&\frac{i}{2}\omega_{\mu}^{ab}\Sigma_{ab},\nonumber\\
\omega_{\mu b}^{a}&=&e^{a}_{\nu}e^{\lambda}_{b}\Gamma^{\nu}_{\mu\lambda}-e^{\lambda}_{b}\partial_{\mu}e^{a}_{\lambda},\nonumber\\
\Sigma_{ab}&=&\frac{i}{4}[\gamma^{a}, \gamma^{b}].\nonumber
\end{eqnarray}
The $\gamma$ matrices satisfy the relationship
\begin{eqnarray}
\{\gamma^{a}, \gamma^{b}\}&=&2\eta^{ab},\nonumber\\
\{\gamma^{\mu}, \gamma^{\nu}\}&=&2g^{\mu\nu}.
\end{eqnarray}
Solving the Eq. (\ref{q1}) and applying the WKB approximation will lead us to the following coupled differential equations
\begin{eqnarray}
0&=&A\bigg(\frac{-\chi}{r}\partial_{\phi}\mathcal{Y}+m+\frac{\beta m^{2}\chi}{r}\partial_{\phi}\mathcal{Y}+
\frac{\beta \chi^{3}f}{r}\partial_{\phi}\mathcal{Y}(\partial_{r}\mathcal{Y})^{2}+\frac{\beta\chi^{3}}{r^{3}}(\partial_{\phi}\mathcal{Y})^{3}
-\beta m^{3}-\beta m\chi^{2}f(\partial_{r}\mathcal{Y})^{2}\nonumber\\
&-&\frac{\beta m\chi^{2}}{r^{2}}(\partial_{\phi}\mathcal{Y})^{2}\bigg)\nonumber\\
&+&B\bigg(-\frac{\chi}{\sqrt{f}}\partial_{t}\mathcal{Y}-\chi\sqrt{f}\partial_{r}\mathcal{Y}
+\beta m^{2}\chi\sqrt{f}\partial_{r}\mathcal{Y}+\beta \chi^{3}f\sqrt{f}(\partial_{r}\mathcal{Y})^{3}
+\frac{\beta\chi^{3}\sqrt{f}}{r^{2}}\partial_{r}\mathcal{Y}(\partial_{\phi}\mathcal{Y})^{2}\bigg),\nonumber \\
 \label{q2} \\
0&=&A\bigg(\frac{\chi}{\sqrt{f}}\partial_{t}\mathcal{Y}-\chi\sqrt{f}\partial_{r}\mathcal{Y}
+\beta m^{2}\chi\sqrt{f}\partial_{r}\mathcal{Y}+\beta \chi^{3}f\sqrt{f}(\partial_{r}\mathcal{Y})^{3}
+\frac{\beta\chi^{3}\sqrt{f}}{r^{2}}\partial_{r}\mathcal{Y}(\partial_{\phi}\mathcal{Y})^{2}\bigg),\nonumber\\
&+&B\bigg(\frac{\chi}{r}\partial_{\phi}\mathcal{Y}+m-\frac{\beta m^{2}\chi}{r}\partial_{\phi}\mathcal{Y}-
\frac{\beta \chi^{3}f}{r}\partial_{\phi}\mathcal{Y}(\partial_{r}\mathcal{Y})^{2}-\frac{\beta\chi^{3}}{r^{3}}(\partial_{\phi}\mathcal{Y})^{3}
-\beta m^{3}-\beta m\chi^{2}f(\partial_{r}\mathcal{Y})^{2}\nonumber\\
&-&\frac{\beta m\chi^{2}}{r^{2}}(\partial_{\phi}\mathcal{Y})^{2}\bigg).\nonumber\\
\label{q3}
\end{eqnarray}
For non-trivial solution of the above coupled equations the determinant of the coefficient matrix must vanish. Thus, for this purpose, ignoring higher powers of $\beta$, we reach at the following constraint
\begin{eqnarray}
0&=&\frac{\chi^{2}}{f}(\partial_{t}\mathcal{Y})^{2}-\chi^{2}f(\partial_{r}\mathcal{Y})^{2}
-\frac{\chi^{2}}{r^{2}}(\partial_{\phi}\mathcal{Y})^{2}+m^{2}\nonumber\\
&+&\beta\bigg(-2m^{4}+\frac{4\chi^{4}f}{r^{2}}(\partial_{r}\mathcal{Y})^{2}(\partial_{\phi}\mathcal{Y})^{2}
+2\chi^{4}f^{2}(\partial_{r}\mathcal{Y})^{4}+\frac{2\chi^{4}}{r^{4}}(\partial_{\phi}\mathcal{Y})^{4}\bigg). 
\label{q4}
\end{eqnarray}
Using Eq. (\ref{2a}) into above equation, we get
\begin{eqnarray}
0&=&\frac{\chi^{2}\omega^{2}}{f}-\chi^{2}f(\partial_{r}W)^{2}
-\frac{\chi^{2}j^{2}}{r^{2}}+m^{2}\nonumber\\
&+&\beta\bigg(-2m^{4}+\frac{4\chi^{4}f j^{2}}{r^{2}}(\partial_{r}W)^{2}
+2\chi^{4}f^{2}(\partial_{r}W)^{4}+\frac{2\chi^{4}j^{4}}{r^{4}}\bigg). 
\label{q5}
\end{eqnarray}
For the radial action we take the form $W=W_{0}+\beta W_{1}$ \cite{beta} and solve the above equation, obtaining
\begin{equation}\label{q6}
W_{\pm}=\pm\frac{\pi i\omega}{f^{\prime}(r_{+})}\bigg(1+\beta\Xi\bigg),
\end{equation}
where
\begin{equation}\label{q7}
\Xi=m^{2}+\frac{f^{\prime}(r_{+})}{8r_{+}^{4}\big[(M-K)\alpha^{2}
-\Lambda\big]^{2}}\left\{4\chi\partial_{r}\chi\omega^{2}r_{+}-2\chi^{2}\omega^{2}r_{+}-m^{2}r_{+}^{2}f^{\prime}(r_{+})
+\chi^{2}j^{2}f^{\prime}(r_{+})\right\}.
\end{equation}
Now similarly proceeding as in the previous section, the tunneling probability of Dirac particles across the horizon is found to be
\begin{equation}\label{q8}
\Gamma = exp \bigg(\frac{4\pi\omega}{f^{\prime}(r_{+})}\big(1+\beta\Xi\big)\bigg),
\end{equation}
this in turns gives the Hawking temperature
\begin{equation}\label{q9}
T_{H} = \frac{f^{\prime}(r_{+})}{4\pi\big(1+\beta\Xi\big)}.
\end{equation}
The quantity outside the parenthesis is the original Hawking temperature that we already obtained in the previous section. The correction term $\Xi$ depends not only on the mass of the black hole but also on the mass, energy and angular momentum of the emitted particles. The original Hawking temperature is recovered in the absence of $\beta.$

\section{Discussion}

In this work, we have calculated the Hawking temperature of Dirac particle's from accelerating BTZ black
hole by using the quantum tunneling approach. We have solved the Dirac equation and obtained Hawking temperature of accelerating BTZ black hole by applying the WKB approximation. The results of this study are consistent with the previous findings \cite{41}, in the absence of the acceleration parameter. Our results show that the temperature is directly proportional to the square of the acceleration parameter.

Tunneling probability of Dirac particles do not depend on the mass of particles, which implies that the emission rate of massive as well as massless particles is same. We have also explored the quantum corrections to the Hawking temperature. We found that, the quantum correction term in the Hawking temperature contains the mass, energy and angular momentum of the emitting particles. Whereas if we ignore the quantum corrections then the Hawking temperature becomes independent of these parameters of the emitting particles. Hawking temperature is calculated from the tunneling probabilities. The difference between classical and semi-classical approach is that, only outgoing particles cross the horizon in the first case and both the ingoing and outgoing particles can cross the horizon in the latter case. For critical values of mass and acceleration, total emission flux in both the cases which we have investigated in this work can vanish, depending on the values of $\hbar$.

\section*{Acknowledgment}
The authors would like to thank the anonymous referee for the valuable comments and suggestions to improve the quality of the paper.

\end{document}